\documentclass[useAMS,usenatbib,twocolumn]{mn2e}
\usepackage{graphicx}
\usepackage{subfigure}

\newcommand{\mpt}{\mathrm{.}}
\newcommand{\mcm}{\mathrm{,}}

\newcommand{\Vec}[1]{ \mbox{\boldmath$ #1 $} }

\newcommand{\apjl}{ApJ}
\newcommand{\apj}{ApJ}
\newcommand{\apjs}{ApJS}
\newcommand{\mnras}{MNRAS}

\title[Strong lens inversion of Cl~0024+1654]
      {Non-parametric strong lens inversion of Cl~0024+1654: 
	   illustrating the monopole degeneracy}
\author[J. Liesenborgs, S. De Rijcke, H. Dejonghe and P. Bekaert]
{J. Liesenborgs$^1$\thanks{Corresponding author:
jori.liesenborgs@uhasselt.be}, S. De Rijcke$^2$\thanks{Postdoctoral
Fellow of the Fund for Scientific Research - Flanders
(Belgium)(F.W.O)}, H. Dejonghe$^2$ and P. Bekaert$^1$\\ $^1$
Expertisecentrum voor Digitale Media, Universiteit Hasselt,
Wetenschapspark 2, B-3590, Diepenbeek, Belgium \\ $^2$ Sterrenkundig
Observatorium, Universiteit Gent, Krijgslaan 281, S9, B-9000, Gent,
Belgium}

\begin{document}
	
\date{} % TODO

\pagerange{\pageref{firstpage}--\pageref{lastpage}} \pubyear{2008}

\maketitle \label{firstpage} 

\begin{abstract} 
The cluster lens Cl~0024+1654 is undoubtedly one of the most beautiful
examples of strong gravitational lensing, providing five large images
of a single source with well-resolved substructure. Using the
information contained in the positions and the shapes of the images,
combined with the null space information, a non-parametric technique
is used to infer the strong lensing mass map of the central region of
this cluster. This yields a strong lensing mass of $1.60 \times
10^{14} M_\odot$ within a 0.5$'$ radius around the cluster center.
This mass distribution is then used as a case study of the monopole
degeneracy, which may be one of the most important degeneracies in
gravitational lensing studies and which is extremely hard to break.
We illustrate the monopole degeneracy by adding circularly symmetric
density distributions with zero total mass to the original mass map
of Cl~0024+1654. These redistribute mass in certain areas of the mass
map without affecting the observed images in any way. We show that the
monopole degeneracy and the mass-sheet degeneracy together lie at the
heart of the discrepancies between different gravitational lens
reconstructions that can be found in the literature for a given
object, and that many images/sources, with an overall high image
density in the lens plane, are required to construct an accurate,
high-resolution mass map based on strong-lensing data.
\end{abstract}


\section{Introduction}

Due to the gravitational deflection of light, a galaxy or
cluster of galaxies can affect the light that we receive
from background sources. On larger scales, this leads to
slight deformations of the shapes of the background sources,
but close to the center of the deflecting object, the
gravitational lens, more elaborate deformations are
possible. When a background source is sufficiently well
aligned with the gravitational lens, this strong lens
effect can even cause multiple images of said source
to appear. One of the most spectacular examples of strong gravitational
lensing can be seen in the cluster lens Cl~0024+1654. Using
recent ACS observations, one can easily see that five well 
resolved images depict a single source, but even before these
five images were identified, it was clear that three arc
segments were caused by a gravitational lens effect 
\citep{1988lsmu.book..513K}. 

This strong lensing information was first used in
\citet{1992ApJ...400...41K}. The authors of this work noted that these
arc segments do not obey the so-called length theorem
\citep{1990LNP...360...16K}, implying that no simple elliptical lens
model can be used. They show that if perturbations by cluster members
are added, the observed arc lengths can indeed be reconstructed. In
\citet{1995ApJ...441...58W}, a more advanced reconstruction technique
was used, consisting of a smooth lens model perturbed by some smaller
galaxies and a non-parametric source model. Whereas previous work
suggested that the main cluster potential was offset from the largest
galaxy, these authors find that these positions, in fact, agree well.

After the first HST images clearly revealed the presence of five
images, more lensing studies followed. The new, well resolved images
were used in \cite{1996ApJ...461L..83C} to study the source itself, a
blue galaxy containing some interesting dark features and a bar-like
structure. \citet{1998ApJ...498L.107T} use the images to find the
parameters describing elaborate lens and source models. Their
algorithm constructs the complete image plane based on a set of source
and lens parameters and compares the result with the HST
observations. They find that the mass distribution is dominated by a
smooth dark matter component with a considerable core radius, centered
at a position near the largest cluster member.

Much of the earlier mass uncertainties originated from the poorly
established source redshift. \citet{2000ApJ...534L..15B} finally
measured a spectroscopic redshift of 1.675 and used this information
in their own inversion. They found that the image positions can be
accurately reproduced using a model which traces the locations of the
brightest cluster members. In \citet{2007ApJ...661..728J}, a
non-parametric method is used to invert the lens, using both strong
and weak-lensing data. In the strong lensing region, the retrieved
mass profile closely resembles the result of
\citet{2000ApJ...534L..15B}, but according to
\citet{2000ApJ...542L...1S}, the associated velocity dispersion is too
high to correspond to the measured value of 1150 km s$^{-1}$
\citep{1999ApJS..122...51D}.

In the present paper, we employ a non-parametric method to infer the
mass map of Cl~0024+1654. This is the first time that only the
information about the images themselves as well as the null space --
i.e. the region where no images are observed -- is used to reconstruct
the mass distribution of this cluster in the strong lensing region. No
information about the positions of cluster members is used. Clearly,
many other possibilities have already been presented in the past, but
it is not our intention to add to the confusion. Instead, the
reconstruction is used to explain how the different previous inversions are
related to each other.

Below, we will first briefly review the non-parametric technique
that is worked out in detail in previous articles. In 
section \ref{sec:inversion} this method is applied to reconstruct
the mass distribution of Cl~0024+1654, and this result is used
in section \ref{sec:monopole} to illustrate the importance of the
monopole degeneracy. The implications of these observations are
discussed in section \ref{sec:conclusion}.

\begin{figure*}
\centering
\includegraphics[width=0.98\textwidth]{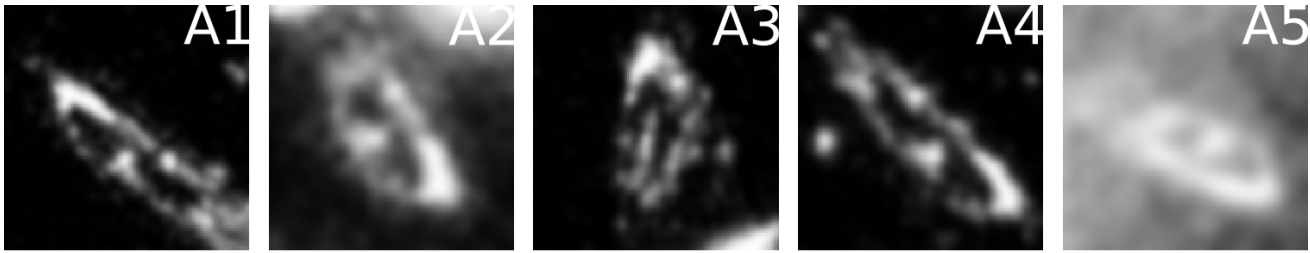}
\caption{The image parts of the five images of source A which
were used in the reconstruction, labeled in the same way as
in the work of \citet{2007ApJ...661..728J}. Due to the extended
nature of these images, several corresponding features 
are easily identified. The images shown here are not
displayed on the same scale.}
\label{fig:inputA}
\end{figure*}

\begin{figure*}
\centering
\subfigure{\includegraphics[width=0.48\textwidth]{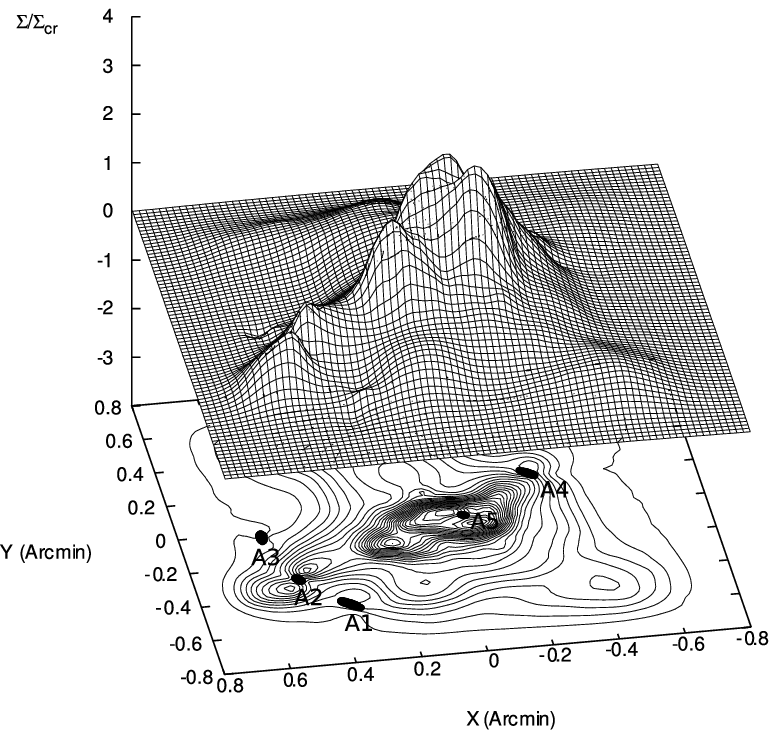}}
\subfigure{\includegraphics[width=0.48\textwidth]{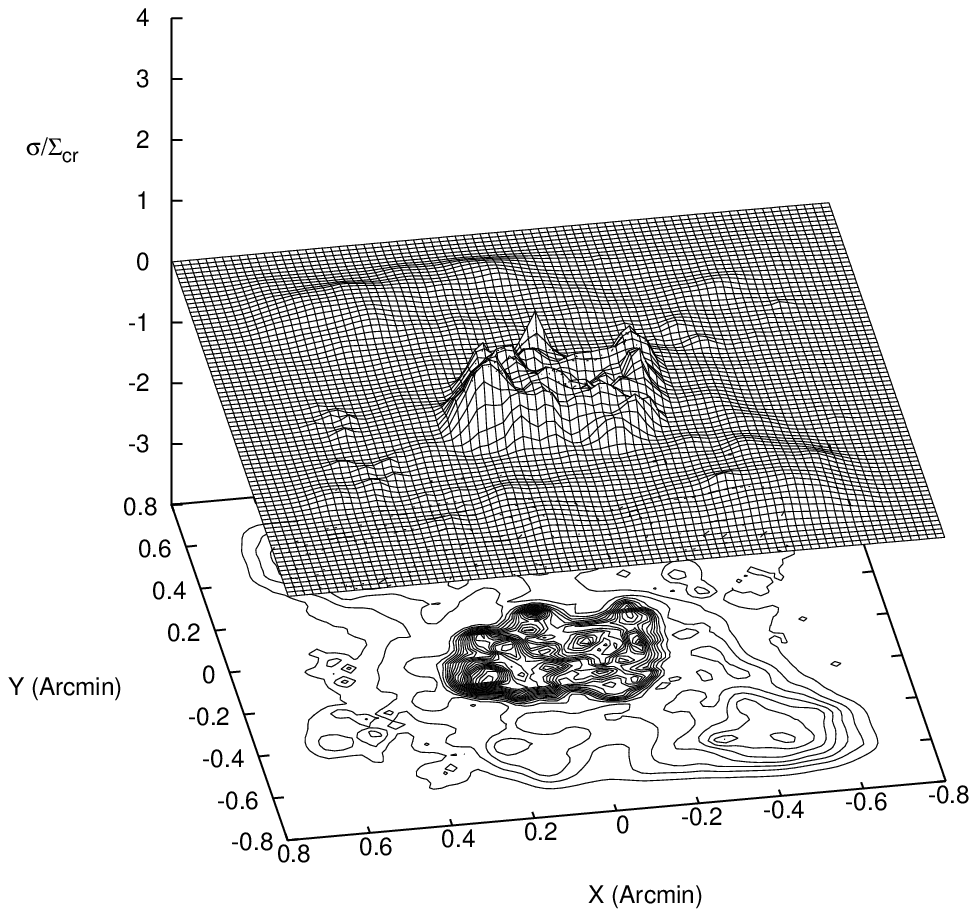}}
\caption{Left panel:~after averaging 28 individual reconstructions, this 
is the resulting mass map for Cl~0024+1654 predicted by our procedure. The
positions of the input images of source A are also indicated in this figure.
The critical density used in this figure corresponds to a redshift $z=3$.
Right panel:~the standard deviation of the individual reconstructions
shows that the different solutions tend to disagree about the exact
shape in the central part of the mass distribution. In particular,
this figure suggests that the mass peak around $(0.2', -0.2')$ in the
left panel should not be regarded as an actual feature.}
\label{fig:sol1}
\end{figure*}

\section{Inversion method}\label{sec:procedure}

Below we shall briefly describe a minor variation of the inversion 
method described in \citet{Liesenborgs} and \citet{Liesenborgs2}.
The interested reader is referred to these works for a detailed
description of the steps involved. A basic knowledge of gravitational
lensing is assumed; we refer to \citet{SchneiderBook} for an
in-depth treatment of the subject.

\subsection{Multi-objective genetic algorithms}

A genetic algorithm is an optimization strategy in which one tries
to produce acceptable solutions to an -- often high-dimensional --
problem using a mechanism inspired by natural selection. In effect,
one tries to breed solutions to a problem.

One starts with a so-called population of genomes, each one encoding
a trial solution to the problem. Based on this population, a new one
is created by combining and mutating existing genomes. It is important
to apply some kind of selection pressure: genomes which are deemed
more fit should have a better chance of creating offspring. If only
one fitness measure is needed, this selection mechanism can be implemented
by first sorting the genomes in a population according to their fitness
and by letting the selection probability depend on the position of the
genome in this sorted population.

A similar approach can be used when more than one fitness measure should
be optimized. One genome is said to dominate another one if it is at least
as good with respect to each fitness criterion and if it is strictly better
regarding at least one criterion. Using this concept of dominance, one can
identify in a population the genomes which are not dominated by any other
genome: the non-dominated set. These genomes should receive the
highest selection probability. If one removes this set from the population,
one can find a new non-dominated set which should receive the 
second-to-highest selection probability, etc.

For more information about both single- and multi-objective genetic 
algorithms, the interested reader is referred to \citet{Deb}.

\subsection{Dynamic grid}

At the start of the inversion procedure, the user is required to specify
a square-shaped area in which the procedure should try to reconstruct the
projected mass distribution. First, this region is subdivided uniformly
into a number of smaller square grid cells, and to each grid cell, a 
projected Plummer sphere \citep{1911MNRAS..71..460P} is associated. The 
widths of these basis functions are proportional to the sizes of the grid 
cells. Based on previous work, we use a Plummer width that is 1.7 times
as large as the size of a cell.

Using a multi-objective genetic algorithm, the inversion procedure then
tries to find weights for these basis functions which are compatible with
the observed gravitational lensing scenario. Once these are found, the
corresponding estimation of the mass distribution is used to create a new
grid, with smaller grid cells in regions containing more mass. Plummer
basis functions are again associated to each grid cell, and the genetic
algorithm will try to find new values for their weights. This refinement
procedure can be repeated a number of times, until an acceptable 
reconstruction is retrieved. Below we explain further which measures are 
used to determine if a solution is acceptable. This dynamic grid system
is also used in \citet{2005MNRAS.362.1247D}.

\subsection{Fitness measures}

If the true mass distribution were known, the corresponding lens
equation would project each image of a single source onto the same
region of the source plane. For this reason, the first fitness
criterion measures the amount of overlap when images are projected
onto their source planes by a trial solution. Each back-projected
image provides an estimate of the shape of the source and determines a
typical scale in the source plane.  By averaging these scales over all
the images of a single source, one obtains a final estimate of a
typical scale for this source. In the original procedure, the
back-projected images were surrounded by rectangles and the distances
between corresponding corners -- measured relative to the estimated
size of the source -- were used to calculate a measure for the overlap
of the images. For the case of Cl~0024+1654, a small variation is
used: because the images are well resolved, several matching points
can be found in the images. The distances -- still relative to the
estimated source size -- between these points when projected onto the
source plane, are then used to calculate a measure for the overlap
between the images.

Not only should the back-projected images form a consistent source, a
candidate solution should also avoid predicting images where none are
observed. A second fitness measure is included to avoid producing
these kinds of solutions. To do so, the region where no images are
observed -- the null space -- is subdivided into a number of small 
triangles. Each triangle is projected back onto the source plane and
is compared to the current estimate of the source shape. For a specific
trial solution, the shape of a source is estimated by projecting all
image points onto the source plane, and by calculating the envelope
of these points. If such a null-space 
triangle and the estimated source shape overlap, this would indicate 
that an extra image is predicted by the current trial solution. Such 
a comparison is done for each null-space triangle and the total amount 
of overlap with the source is used to calculate a fitness measure.

In the case of non-merging images, one would like to avoid critical lines
intersecting the images. To do so, a third fitness criterion is used.
It is easy to detect if a critical line intersects an image: one simply
needs to calculate the sign of the magnification at each image point.
Only if this is the same for all image points, no critical line intersects
the image. In practice, the magnification signs of neighbouring points 
are compared and the fitness measure simply counts the number of pairs
for which the signs change.

\begin{figure*}
\centering
\includegraphics[width=0.98\textwidth]{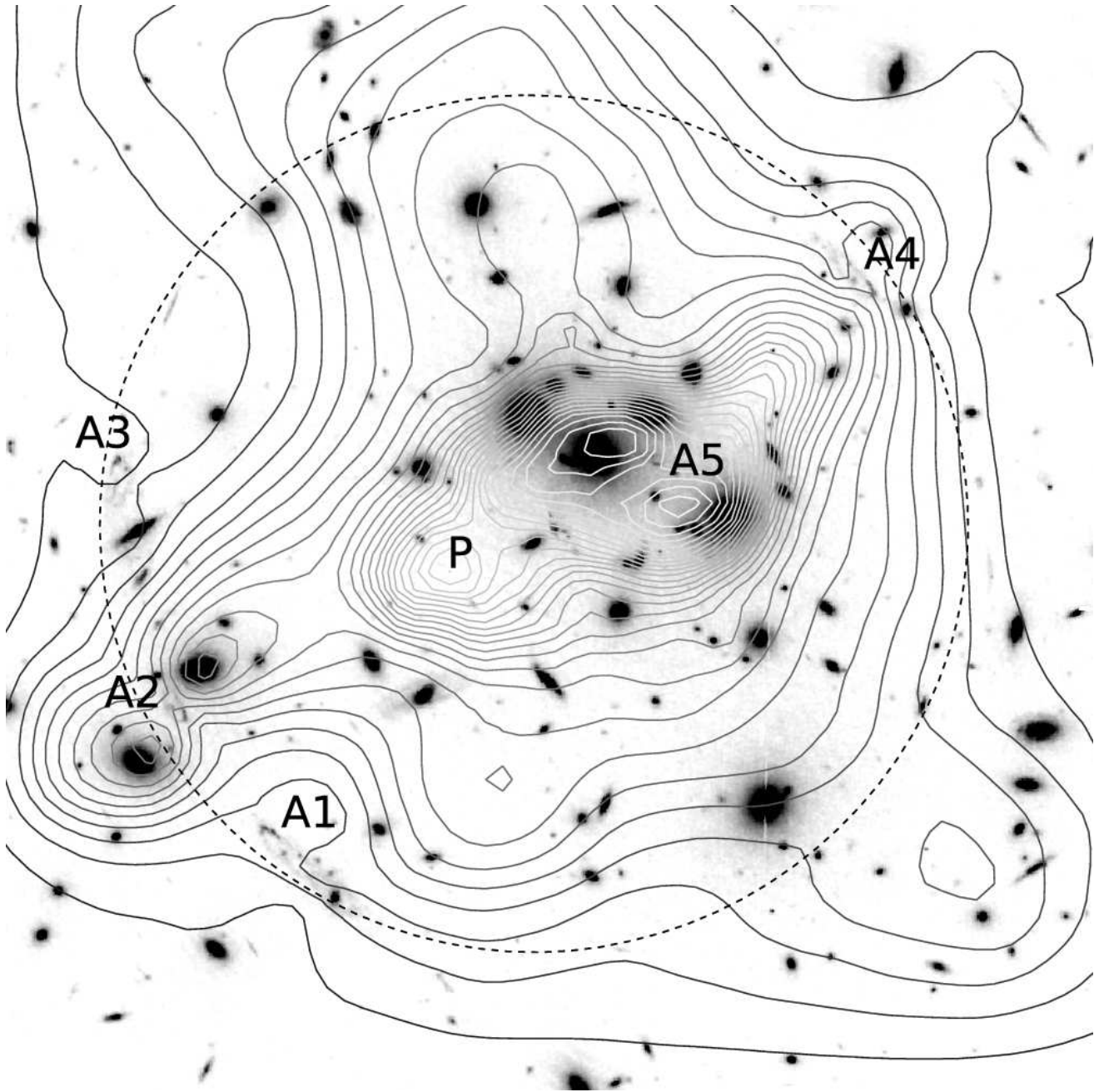}
\caption{The contours of the retrieved mass map in the left panel of 
Fig.~\ref{fig:sol1} are shown on top of the ACS image of the
central cluster region (north is up, east is left). 
The lensing mass is found to be concentrated around the
largest cluster galaxy and the central image A5
is found to be located between two mass peaks, which also
resembles the observed situation. The positions of the two
galaxies in the south-east region are retrieved very
accurately as well. Note that this image also suggests that there
is a density peak labeled P in a region where very few cluster light
originates. The total mass in the region bounded by the dotted
line is found to be $1.60 \times 10^{14} M_\odot$. The area displayed in
this figure is approximately 1.3$'$ $\times$ 1.3$'$.}
\label{fig:sol1overlay}
\end{figure*}

\begin{figure*}
\centering
\includegraphics[width=0.98\textwidth]{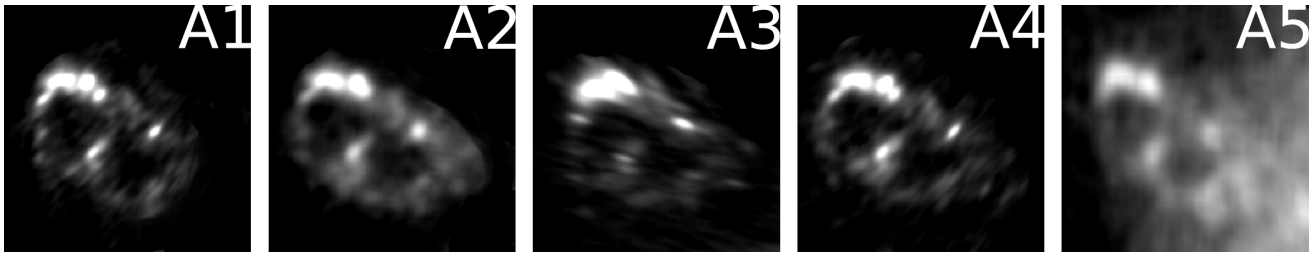}
\caption{When the images shown in Fig.~\ref{fig:inputA} are projected
back onto the source plane, these source shapes are
retrieved. Each figure shows the same region in the source
plane, approximately 3$''$ by 3$''$ in size.}
\label{fig:sourceA}
\end{figure*}

\begin{figure*}
\centering
\subfigure{\includegraphics[width=0.48\textwidth]{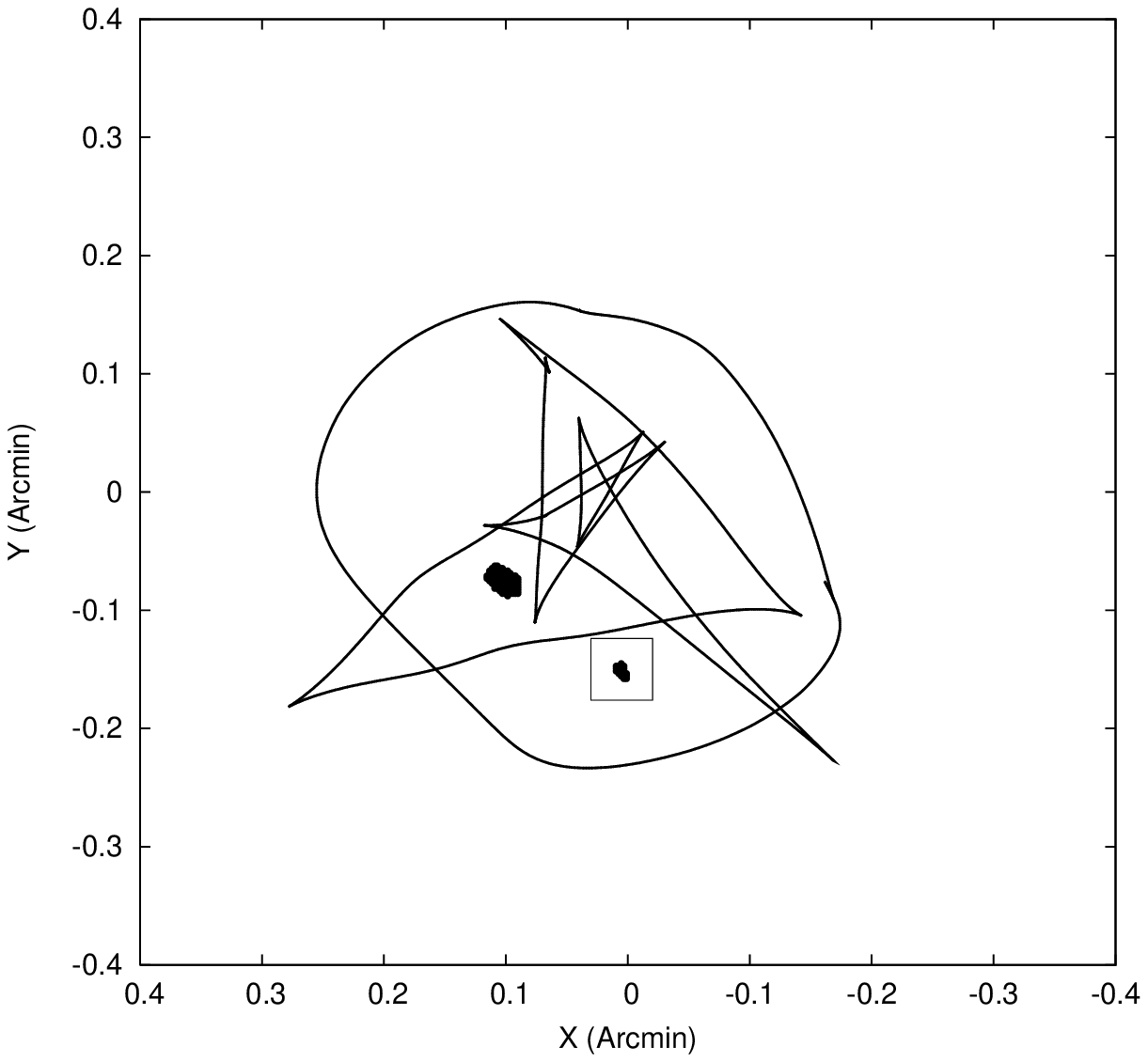}}
\subfigure{\includegraphics[width=0.48\textwidth]{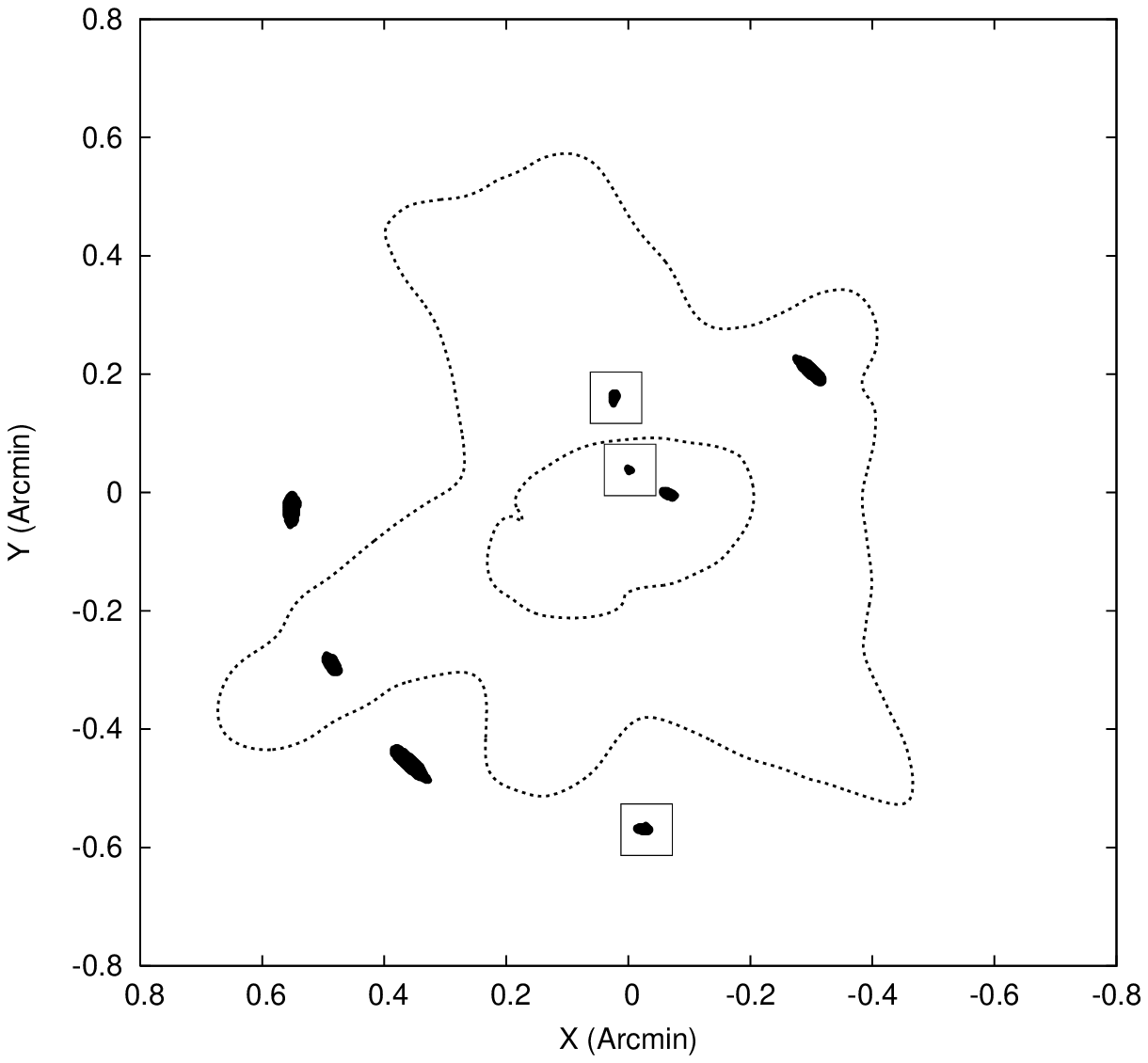}}
\caption{Left panel:~this figure shows the predicted position of 
source A and B as well as the caustics corresponding to the redshift 
of source A. Source B is enclosed by a small square.
Right panel:~when the sources of the left panel are
projected onto their image planes, these image positions arise. 
The image of source B which is closest to the origin was not
part of the input; the model predicts an image at this location.
Other than this image, no additional images were predicted.}
\label{fig:sol1lines}
\end{figure*}

\subsection{Finalizing and averaging}

The dynamic grid method has one disadvantage: regions containing only
a relatively small portion of the total mass will not be subdivided
into smaller grid cells. As a result, the basis functions in such regions
may lack the resolution needed for an accurate reconstruction. To 
overcome this problem, a finalizing step was added to the procedure.
A uniform grid of 64 by 64 grid cells was created and the associated basis
functions are used as small corrections to the current estimate of the 
mass distribution. Because they are corrections, the weights of the basis
functions are allowed to be negative. The genetic algorithm again 
determines appropriate values for these weights.

To create a single candidate solution, first the dynamic grid method is
used to create a first good estimate of the mass distribution and 
afterwards, small-scale corrections are introduced in the finalizing step. 
This entire procedure is then repeated a number of times, each time 
yielding a somewhat different mass distribution. One can then calculate 
the average mass distribution to inspect the features which are common 
in all reconstructions, and one can calculate the standard deviation to 
check in which regions the individual solutions disagree.

\section{Inverting Cl~0024+1654}\label{sec:inversion}

\subsection{Input}

The inversion procedure described above, was applied to the cluster
lens Cl~0024+1654. We use the images of sources A and B as described
in \citet{2007ApJ...661..728J}, at redshifts of 1.675 and 1.3
respectively.  The redshift of the lens itself is 0.395 and angular
diameter distances were calculated in a flat cosmological model with
$H_0 = 71$ km s$^{-1}$ Mpc$^{-1}$, $\Omega_m = 0.27$ and
$\Omega_\Lambda = 0.73$.  The inversion procedure constructs the
lensing mass distribution in a square shaped area of 1.3$'$ by 1.3$'$,
centered on the brightest cluster galaxy. To avoid predicting images
which are located relatively far away, the null space grid measured
3$'$ by 3$'$, centered on the same galaxy. Initially, a uniform grid
of 15$\times$15 is used to place the Plummer basis functions on, and
the grid is refined until approximately 800 basis functions are used.
After this, the finalizing step is executed on a uniform 64$\times$64
grid. Below we shall see that this leads to a very good source
reconstruction.

The gravitational lens creates several large images of source A, a
blue galaxy. A part of the source is mapped onto five easily
identifiable sub-images, as can be seen in Fig.~\ref{fig:inputA}. The
high resolution ACS images allow several corresponding features to be
identified (up to twelve features in some images), which will be used 
to calculate how well the back-projected images overlap.
It is assumed that no other images of
the source are present, so that only the five images themselves are
excluded from the null space for this source. Source B only has two
images, the third one most likely being occluded by the central
cluster members. The complete images are used to estimate the source
size, but only a single point in each image is used to measure how
well the images overlap when projected back onto the source plane
(measured relative to the estimated size of the source). In this case,
not only the images themselves were excluded from the null space, but
also the region in which the central cluster members reside. This
allows the algorithm to predict an unobserved third image anywhere in
that region. For both sources it is assumed that no critical lines
intersect the images which are used.

\subsection{Results}

Using the inversion procedure described in section
\ref{sec:procedure}, the mass map shown in the left panel of
Fig.~\ref{fig:sol1} was obtained after averaging 28 individual
solutions. This number is dictated by the computer time it takes to
generate the individual solutions and by the fact that after averaging
together 15 solutions or more, the average solution does not change
significantly. The largest fraction of the rather steep mass
distribution coincides with the position of the central cluster
members. The central image of source A is located between two density
peaks, which resembles the situation shown in the ACS images. These
facts can be clearly seen when the retrieved mass contours are drawn
on top of the observed situation, as is shown in
Fig.~\ref{fig:sol1overlay}.  This same figures also illustrates the
remarkable accuracy with which the two cluster galaxies enclosing the
image at $(0.5', -0.3')$ are retrieved. We would like to stress again
that these were retrieved automatically; no prior information about
the presence of these galaxies was used. It is these galaxies that
cause the middle image of the three arc segments to be compressed,
thereby causing the violation of the length theorem. The mass inside a
circular region of radius 0.5$'$, centered on $(0.075', -0.075')$ is
found to be $1.60 \times 10^{14} M_\odot$. This region is enclosed by
a dotted line in Fig.~\ref{fig:sol1overlay}.

When the input images of source A are projected back onto the source
plane, a consistent source is produced, as can be clearly seen in
Fig.~\ref{fig:sourceA}. The size of the source is approximately
2.5$''$ (corresponding to 21 kpc). This is larger than both the value
of 1$''$ mentioned in \citet{1996ApJ...461L..83C} and the value of
0.5$''$ mentioned in \citet{2007ApJ...661..728J}, but the general
appearance does agree very well with both works. We shall come back to
this size difference later in section \ref{sec:conclusion}. The
retrieved source positions and caustics at $z=1.675$ are depicted in
the left panel of Fig.~\ref{fig:sol1lines}. If these sources are used
to calculate the image positions, the results shown in the right panel
of Fig.~\ref{fig:sol1lines} are retrieved.  From this image it is
clear that the multi-objective genetic algorithm succeeded in
generating solutions which only predict one extra image (for source B)
and which do not have critical lines intersecting the input images.

\begin{figure*}
\centering
\subfigure{\includegraphics[width=0.48\textwidth]{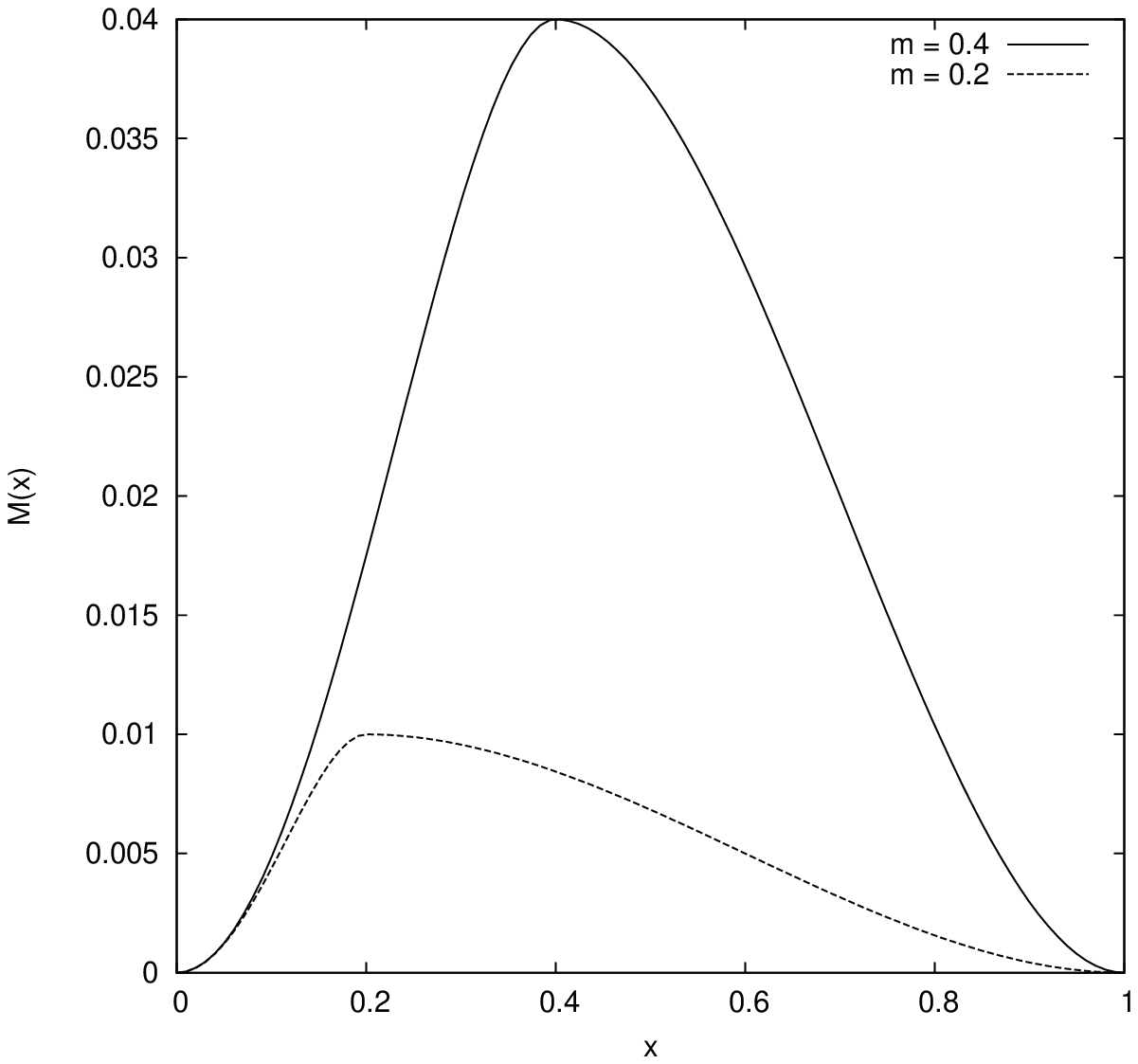}}
\subfigure{\includegraphics[width=0.48\textwidth]{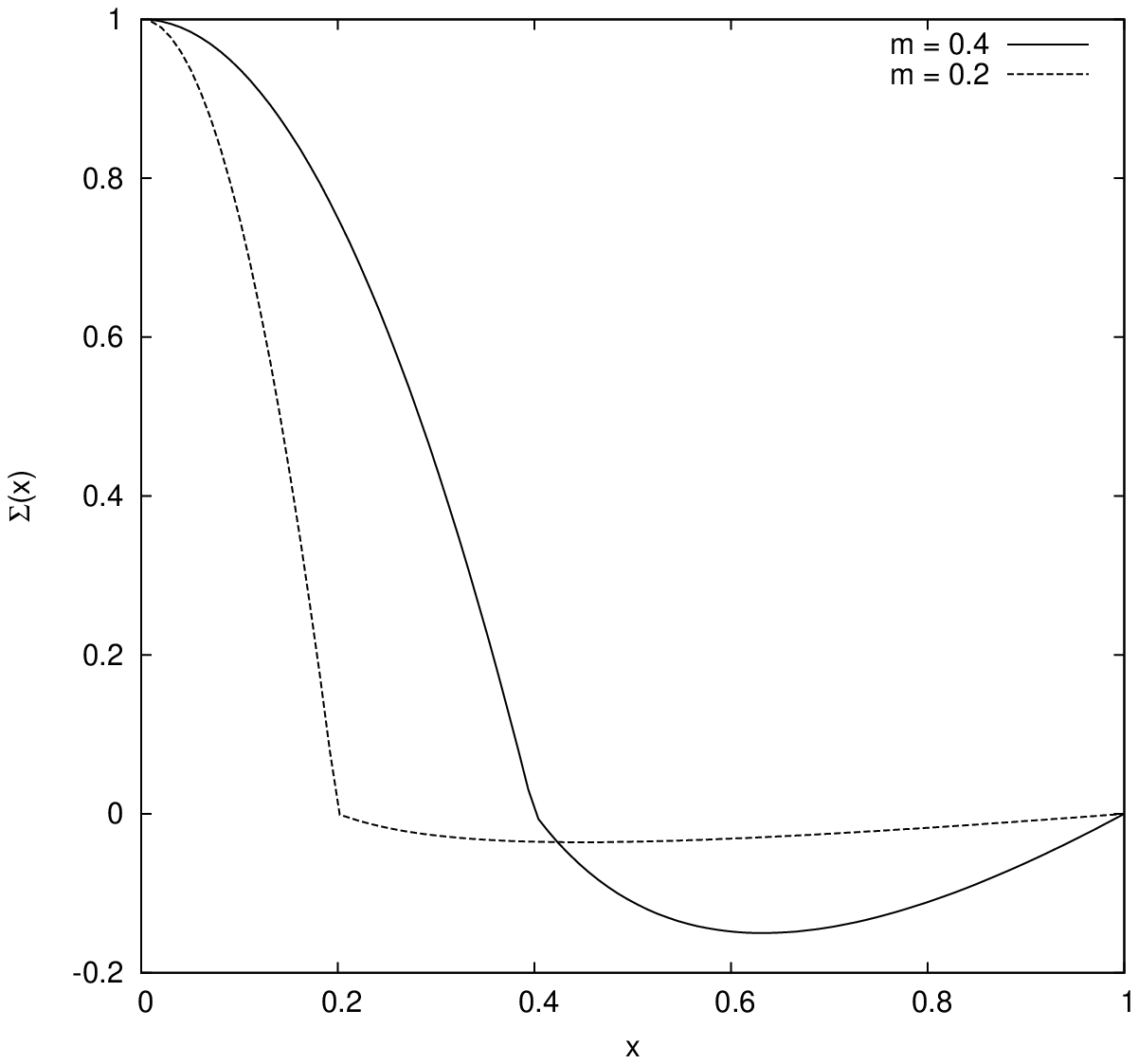}}
\caption{Left panel:~shape of the total mass map of the circularly
symmetric basis functions used to construct degenerate
solutions (see text). The value of $m$ determines the position
of the maximum.
Right panel:~the total mass profiles shown in the left panel
give rise to these density profiles. As the value of $m$ 
becomes smaller, the amplitude of the negative density part
decreases. }
\label{fig:monobasis}
\end{figure*}

On closer inspection of the resulting mass map in Fig.~\ref{fig:sol1}, 
there seems to be an intriguing feature at $(0.2', -0.2')$. At this 
location the mass map shows a clear peak, but in the ACS images no 
cluster member can be seen at this location (Fig.~\ref{fig:sol1overlay}). 
Could this be evidence of dark matter in this cluster? Inspecting the 
standard deviation of the individual solutions helps to shed some light 
on this matter. As can be seen in the right panel of Fig.~\ref{fig:sol1}, 
the individual solutions do not agree well on the exact shape of the
central part of the mass distribution. In fact, the largest 
uncertainty is located precisely around the position of this 
mysterious peak, which suggests that we should be very careful
when trying to interpret this feature.

\section{The monopole degeneracy}\label{sec:monopole}

To verify if any particular feature can be regarded as a real feature 
of the mass map, a question one can ask is the following: can this 
feature be removed from the reconstruction while still obtaining a good 
inversion, given the available constraints? Below, we shall describe how
the monopole degeneracy can help to answer this question and we shall 
apply it to the case of Cl~0024+1654.

For a circularly symmetric mass distribution $\Sigma(\theta)$, the
expression for the bending angle in the lens plane simplifies to
\begin{equation}\label{eq:alphasymm}
\Vec{\hat{\alpha}}(\Vec{\theta}) = \frac{4 G M(\theta)}
{c^2 D_{\rm d} \theta^2}\Vec{\theta}\mcm
\end{equation}
in which $M(\theta)$ is the total mass enclosed within an angle
$\theta$ from the center of this mass distribution:
\begin{equation}\label{eq:enclosedmass}
M(\theta) = 2\pi D_{\rm d}^2\int_0^\theta\Sigma(\theta')\theta' d\theta'\mcm
\end{equation}
and $D_{\rm d}$ is the angular diameter distance between lens and
observer.  From equation (\ref{eq:alphasymm}), it is clear that a
circularly symmetric mass distribution of which the total mass is zero
beyond a specific radius, does not produce a gravitational lens effect
outside said radius. If such a mass distribution is added to an
existing one, the original lens equation will be modified only inside
the circular region in which it has non-zero mass.

Consider a lens mass map $M(x)$ specified by $M_A(x)$ in $[0,m]$, by
$M_B(x)$ in $[m,1]$ and which is zero beyond the unit radius:
\begin{eqnarray}
M_A(x) & = & -\frac{1}{4 m^2}x^4 + \frac{1}{2}x^2 \\
M_B(x) & = & \frac{m^2}{4(m-1)^3} \times \nonumber \\
&   & \left[-2 x^3 + 3(m+1) x^2 -6 m x + 3 m - 1\right]\mpt
\end{eqnarray}
The shape of such a function is shown in the left panel of 
Fig.~\ref{fig:monobasis} for two different values of $m$, which specifies
the position of the maximum. The right panel of the same figure shows
the associated density profiles, which are composed of two parts as well:
\begin{eqnarray}
\Sigma_A(x) & = & \frac{1}{x}\frac{d M_A}{d x} = -\frac{1}{m^2}x^2+1 \\
\Sigma_B(x) & = & \frac{1}{x}\frac{d M_B}{d x} \nonumber \\
& = & \frac{m^2}{4(m-1)^3}\left[-6 x + 6(m+1) -\frac{6 m}{x} \right] \mpt
\end{eqnarray}
Clearly, the smaller the value of $m$, the flatter the density profile
becomes after this point. Using such a profile, it is possible to
introduce or erase a peak in an existing mass map without changing
much to the rest of the distribution and while conserving the total
mass.

Using basis functions of this type, we can build a more complex mass
distribution that, when added to an existing gravitational lens
reconstruction, will produce an equally acceptable solution. To do so,
the region of interest is subdivided into a number of square-shaped
grid cells. For each grid cell, the distance from its center to the
nearest image is calculated. If this distance is relatively large
compared to the size of the grid cell -- e.g. at least four times as
large -- a basis function is associated to this cell. The distance to
the nearest image is used as the unit length; the width of the
non-negative part $\Sigma_A$ is set proportional to the size of the
grid cell.  This implies that for a specific basis function, all the
images lie in the area within which the total mass of the basis
function is zero. Since the lens equation for a circularly symmetric
basis function only depends on the total mass within a specific
radius, in this case the lens equation at the position of the images
will be unaffected when such a basis function is added to the existing
mass distribution. The property that the total mass of the basis
function outside a certain radius is zero, is clearly an essential
feature in this approach. Similarly, when all the basis functions on
the grid are considered, the lens equation at the location of the
images will not be influenced, independent of the precise weight
values of the basis functions. Everywhere else, the lens equation will
indeed be modified, meaning that extra images may be predicted,
depending on the precise weight values used.

\begin{figure*}
\centering
\subfigure{\includegraphics[width=0.48\textwidth]{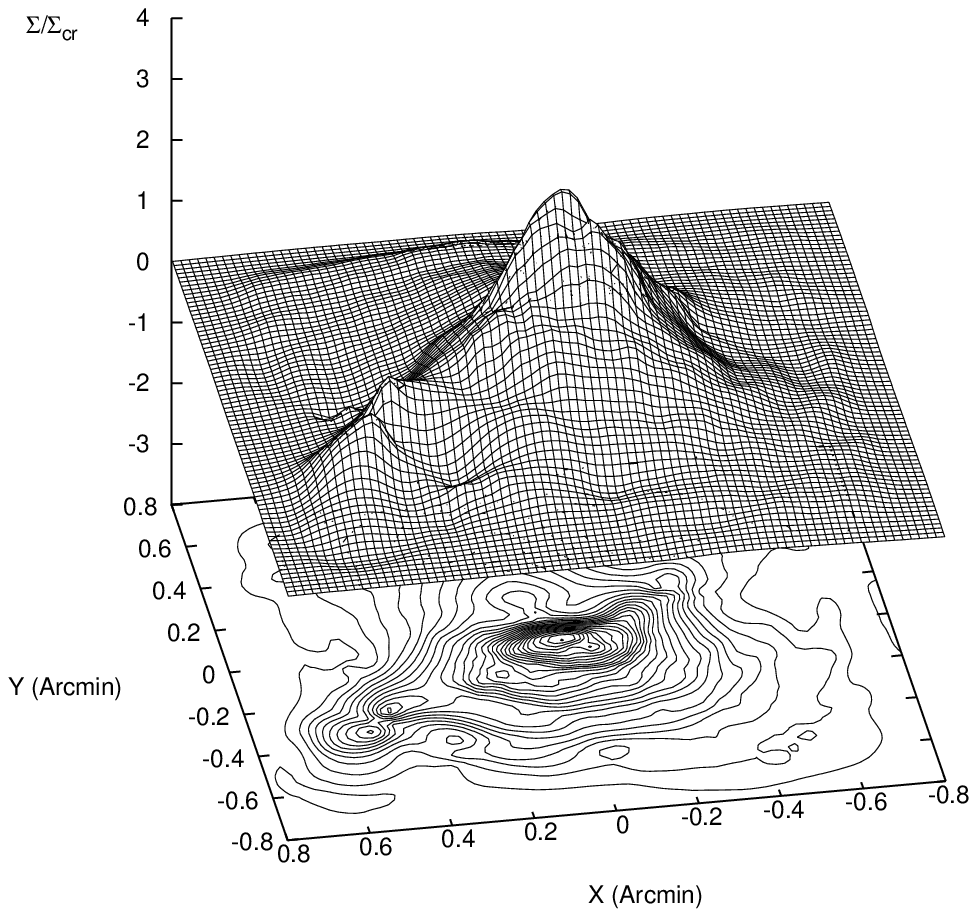}}
\subfigure{\includegraphics[width=0.48\textwidth]{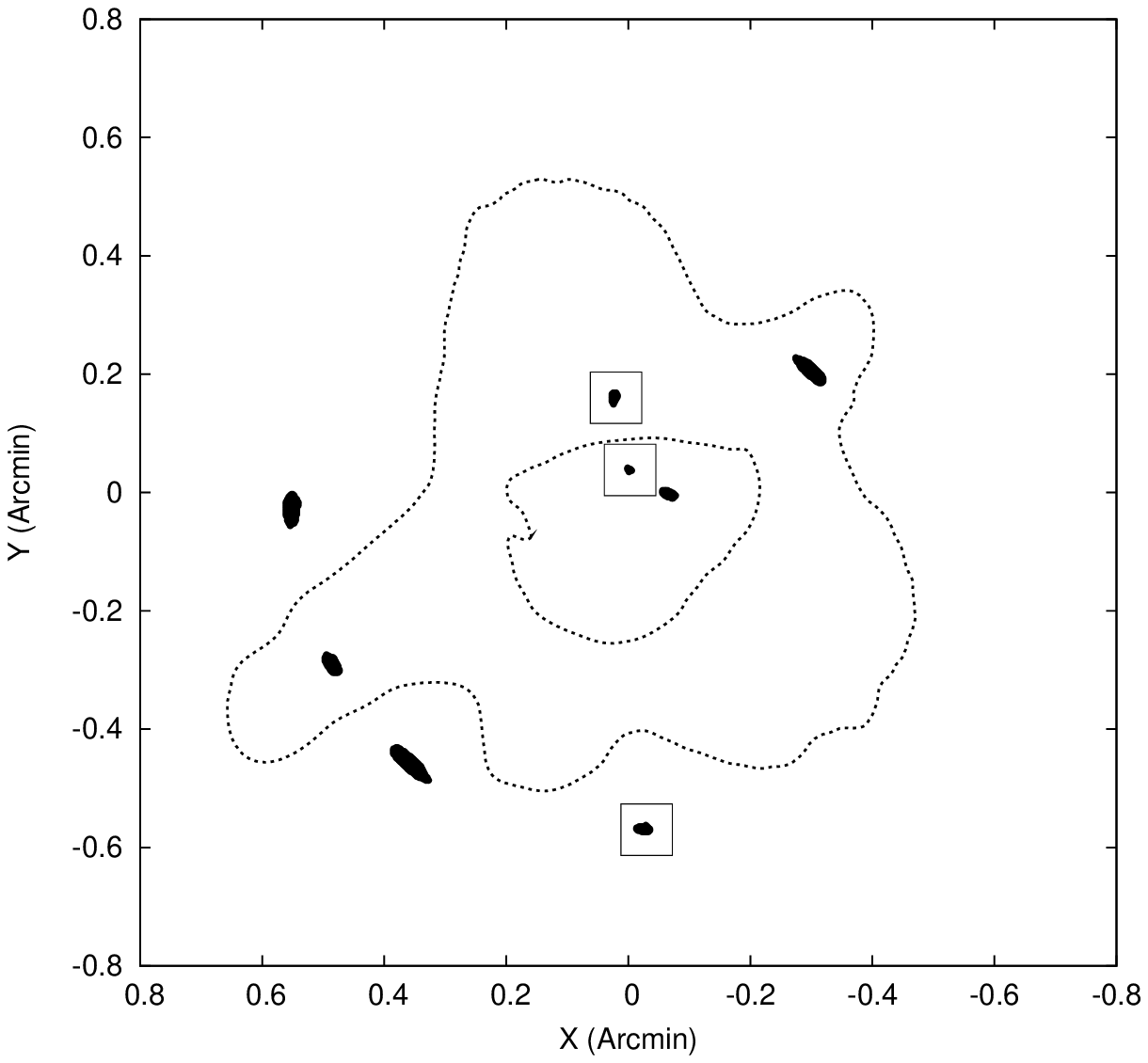}}
\caption{Left panel:~if the grid based method to redistribute mass
is applied to the mass map shown in the left panel of Fig.~\ref{fig:sol1},
this new distribution is obtained. The peak at $(0.2', -0.2')$
has automatically been removed and the overall distribution
has become less steep.
Right panel:~the mass distribution in the left panel predicts
the images shown in this figure, which are indistinguishable
from the images in Fig.~\ref{fig:sol1lines} (right panel). The
critical lines on the other hand, do display some changes, reflecting
the modifications to the lens equation.}
\label{fig:degen}
\end{figure*}

In the case of Cl~0024+1654, it then becomes immediately clear that
the peak at $(0.2', -0.2')$ can easily be removed by creating a
degenerate solution. Even by adding a single basis function with an
appropriate width and height to the existing solution, the feature can
be eliminated.  It can also automatically be removed using the
grid-based procedure described above, as can be seen in the left panel
of Fig.~\ref{fig:degen}.  In this example, a 32 by 32 grid was used,
and the weights were determined by a genetic algorithm. The goal of
the optimization was to keep the gradient of the resulting mass map as
low as possible. To obtain a smooth result, the procedure was repeated
for twenty of such grids, each with a small random offset. As can be
seen in the figure, this does not only remove the peak at $(0.2', -0.2')$, 
but also reduces the overall steepness. Also note that one of
the peaks between which the central image of source A originally
resided, has been erased almost entirely. The resulting mass map,
consisting of one smooth component and two perturbing components, at
least qualitatively resembles the models used by
\citet{1992ApJ...400...41K} and \citet{1995ApJ...441...58W}. The right
panel of Fig.~\ref{fig:degen} shows the critical lines at the redshift
of source A, as well as the images predicted by the new solution.
Because this newly created solution does not modify the lens equation
in the regions of the images and because no extra images are created,
the fitness values are exactly the same as those of the solution in
Fig.~\ref{fig:sol1}. For this reason, both mass maps are equally
acceptable solutions.

\section{Discussion and conclusion}\label{sec:conclusion}

In this article we have applied a previously described non-parametric
inversion method to the cluster lens Cl~0024+1654. The method uses
both the information from the extended images and the null space and 
can easily be adapted to incorporate other available constraints. 
It requires
the user to specify a square shaped area in which the algorithm should
search for the mass distribution and it is assumed that no mass resides
beyond the boundaries of this region, but no other bias is present.
Different runs of the inversion method can produce results that differ
somewhat, depending on the amount of constraints available. This allows
one to inspect which features are common in all inversions and which
aspects tend to differ.

Using this inversion procedure we obtained an averaged mass map which
clearly displays features that can also be seen in the ACS images. The
most recent gravitational lensing study of this lens, is that of
\citet{2007ApJ...661..728J}, which used both strong and weak lensing
data. The strong lensing mass of $1.60\times 10^{14} M_\odot$ is less
than the value of $(1.79 \pm 0.13)\times 10^{14} M_\odot$ found in
their study, but it is still in good agreement. We mentioned earlier
that our size estimate for source A is higher than found in other
works. This is a well-known consequence of a generalized version of
the mass-sheet degeneracy, for which the name steepness
degeneracy is more appropriate. As we showed in \citet{Liesenborgs3}, 
this steepness degeneracy is very hard to break for lensing systems 
with only a handful of sources, even if these have different 
redshifts. As the original mass-sheet degeneracy, the generalized
degeneracy leaves the observed images identical but the reconstructed 
sources are scaled versions of the original ones while the density 
profile of the lens becomes less steep.

\begin{figure}
\centering
\includegraphics[width=0.48\textwidth]{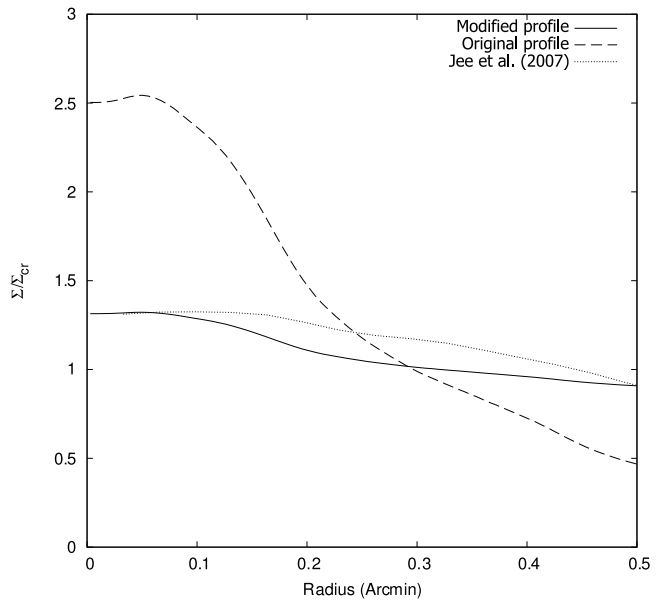}
\caption{The density profile in the circular region indicated in
Fig.~\ref{fig:sol1overlay} is
described by the dotted curve. If this profile is
scaled down by a factor of five and a mass sheet is
added to keep the strong lensing mass constant, the
the profile described by the thick black line is
obtained. In the strong lensing region,
it clearly resembles the profile shown in the work 
of \citet{2007ApJ...661..728J}, suggesting that the
results shown here differ mainly by the mass-sheet
degeneracy. This figure also clearly shows that the
strong lensing mass estimate from this work differs
from the one in \citet{2007ApJ...661..728J}.}
\label{fig:sol1profiledegen}
\end{figure}

The relation with the inversion of \citet{2007ApJ...661..728J} can be
revealed by comparing the predicted source sizes. The size of source A
in our inversion is five times larger than in their work, thereby
identifying the scale factor in the mass-sheet degeneracy.  When we
downscale our mass reconstruction by a factor of five and add a
constant sheet of mass in such a way that the strong lensing mass is
unaffected, the circularly averaged density profile in
Fig.~\ref{fig:sol1profiledegen} is obtained (thick black line). This
clearly shows much resemblance to the profile found in
\citet{2007ApJ...661..728J} in the strong lensing region. Note that
since our reconstruction procedure only looks for mass in a region
which is 1.3$'$ $\times$ 1.3$'$ in size, the profile will quickly drop
to zero beyond the range shown in the figure.

\begin{figure}
\centering
\includegraphics[width=0.48\textwidth]{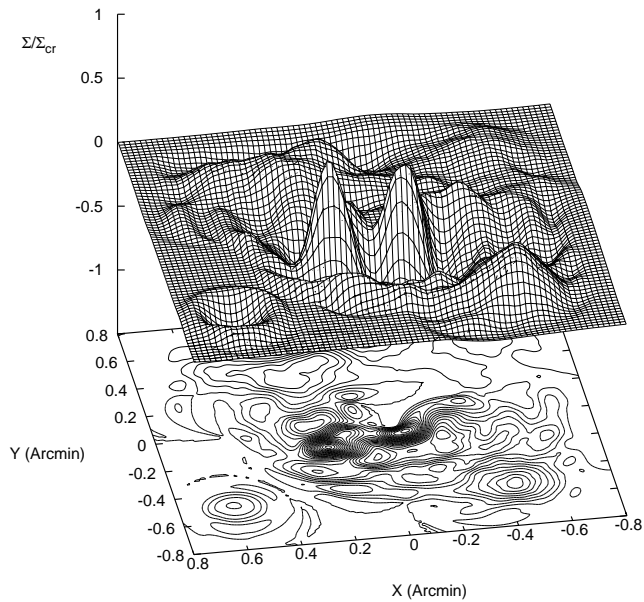}
\caption{This plot shows the differences between the mass distributions
in Fig.~\ref{fig:sol1} and Fig.~\ref{fig:degen}. Clearly,
the structure has been altered in a way which does not display
any particular symmetry.}
\label{fig:degendiff}
\end{figure}

When the monopole degeneracy was applied to the case of Cl~0024+1654,
a simple optimization routine was used to remove substructure from the
previously obtained mass distribution. However, there is no general rule
as to how the mass map may be modified. For example, with some extra 
effort the existing mass map could have been transformed into one which 
followed the light more closely, or which corresponded better to the
available X-ray data \citep{2004ApJ...601..120O}. The only constraints 
which matter in this respect are the absence of unobserved images and 
possibly dynamic measurements. Image positions, fluxes and time delays 
are completely unaffected by this type of degeneracy, which allows you 
to redistribute matter in any number of ways. This freedom is illustrated
in Fig.~\ref{fig:degendiff}, which depicts the differences between the
two mass distributions shown in this article. 

The monopole degeneracy seems to be under-appreciated: the only direct
application that can be found is in the work of
\citet{2006ChJAA...6..141Z}, where circularly symmetric modifications
of power-law models for PG~1115+080 were explored. Yet from the
discussion above it is clear that the degeneracy is an important
aspect of any gravitational lens inversion, as it can be used to
introduce or remove many kinds of features. The explanation in section
\ref{sec:monopole} illustrates the importance of the distance between
the images, implying that the resolution that can be obtained when
inverting a strong gravitational lens system is determined by the
local density of the images. This fact is also mentioned in the work
of \citet{LensPerfect}, but was not linked to the monopole degeneracy.
It is also interesting to note that when the total mass in the region
indicated in Fig.~\ref{fig:sol1overlay} is calculated for the
degenerate solution, one finds the slightly larger value of $1.62
\times 10^{14} M_\odot$.  This indicates how this degeneracy may be
responsible for differences in strong lensing masses in different
studies.

Using both the generalized mass-sheet degeneracy and the monopole
degeneracy as described in this work, it seems likely that the
majority of differences between existing lens models can be
explained. The first thing that one needs to do, is look at the
predicted source sizes. This readily identifies the presence of the
mass-sheet degeneracy. When this is compensated for, the remaining
differences can then be minimized by redistributing the mass using the
monopole degeneracy (which can also easily alter the steepness of the
mass distribution). Clearly, if an accurate mass map is required
without additional assumptions about the shape of the distribution, a
large amount of images are needed. Without sufficient coverage by
images, a fundamental and large uncertainty exists in the regions
between the images. This can only be resolved by identifying
additional multiply imaged sources, and not by more detailed
observations of the existing images (although this can reveal
small-scale substructure in the vicinity of these images). This
fundamental uncertainty in the overall lens equation also implies that
care must be taken when using an existing lens model in trying to
identify new multiply imaged sources.

\section*{Acknowledgment}

The Cl~0024+1654 image data presented in this paper were obtained from the 
Multimission Archive at the Space Telescope Science Institute (MAST). 
STScI is operated by the Association of Universities for Research in 
Astronomy, Inc., under NASA contract NAS5-26555. Support for MAST for 
non-HST data is provided by the NASA Office of Space Science via grant 
NAG5-7584 and by other grants and contracts.

\bsp 
\label{lastpage}

\end{document}